\documentclass[fleqn,usenatbib]{mnras}

\usepackage{newtxtext,newtxmath}

\usepackage[T1]{fontenc}
\usepackage{ae,aecompl}

\usepackage{graphicx}	
\usepackage{amsmath}	
\usepackage{amssymb}	

\usepackage{color}

\usepackage{color}

\newcommand{\oiii}{[O{\sc\,iii}]}

\title[Line confusion and BAO shift]{Line confusion in  spectroscopic surveys and its possible effects: Shifts in Baryon Acoustic Oscillations position}

\author[E. Massara et al.]{
Elena Massara$^{1,2,3}$\thanks{E-mail: elena.massara.cosmo@gmail.com}, Shirley Ho$^{2}$, Christopher M. Hirata$^{4}$ , Joseph DeRose$^{3,5,6}$, 
\newauthor
Risa H. Wechsler$^{7,8,9}$, Xiao Fang$^{10}$  
\\
$^{1}$Waterloo Centre for Astrophysics, University of Waterloo, 200 University Ave W, Waterloo, ON N2L 3G1, Canada \\
$^{2}$Center for Computational Astrophysics, Flatiron Institute, 162 5th Avenue, New York, NY 10010 USA \\
$^{3}$Berkeley Center for Cosmological Physics, University of California, Berkeley, CA 94720, USA\\
$^{4}$Center for Cosmology and AstroParticle Physics, Department of Physics, The Ohio State University, 191 W Woodruff Ave,\\ Columbus OH 43210, USA\\
$^{5}$Santa Cruz Institute for Particle Physics, Santa Cruz, CA 95064, USA\\
$^{6}$Lawrence Berkeley National Laboratory, 1 Cyclotron Road, Berkeley, CA 93720\\
$^{7}$Kavli Institute for Particle Astrophysics \& Cosmology, P. O. Box 2450, Stanford University, Stanford, CA 94305, USA\\
$^{8}$Department of Particle Physics and Astrophysics, SLAC National Accelerator Laboratory, Stanford, CA 94305, USA\\
$^{9}$Department of Physics, Stanford University, 382 Via Pueblo Mall, Stanford, CA 94305, USA\\
$^{10}$Department of Astronomy and Steward Observatory, University of Arizona, 933 N Cherry Ave, Tucson, AZ, 85719, USA\\
}

\date{Accepted XXX. Received YYY; in original form ZZZ}

\pubyear{2020}

\begin{document}
\label{firstpage}
\pagerange{\pageref{firstpage}--\pageref{lastpage}}
\maketitle

\begin{abstract}
Roman Space Telescope will survey about 17 million emission-line galaxies over a range of redshifts. Its main targets are H$\alpha$ emission-line galaxies at low redshifts and \oiii\ emission-line galaxies at high redshifts. The Roman Space Telescope will estimate the redshift these galaxies with single line identification. This suggests that other emission-line galaxies may be misidentified as the main targets. In particular, it is hard to distinguish between the H$\beta$ and \oiii\ lines as the two lines are close in wavelength and hence the photometric information may not be sufficient to separate them reliably. 
Misidentifying H$\beta$ emitter as \oiii\ emitter will cause a shift in the inferred radial position of the galaxy by approximately 90 Mpc/h. This length scale is similar to the Baryon Acoustic Oscillation (BAO) scale and could shift and broaden the BAO peak, possibly introduce errors in determining the BAO peak position. We qualitatively describe the effect of this new systematic and further quantify it with a lightcone simulation with emission-line galaxies. 
\end{abstract}




\section{Introduction}

The next generation of spectroscopic large-scale structure surveys, such as the Wide Field Infrared Spectroscopic Telescope (hereafter, Roman Space Telescope\footnote{\url{https://roman.gsfc.nasa.gov/}}) and the Euclid mission\footnote{\url{https://www.euclid-ec.org}}, will be able to observe a large fraction of the sky significantly deeper than previously possible. Larger and deeper maps will allow us to obtain more precise measurements of the expansion history of the Universe and of the growth of structure, leading to a better understanding of the accelerating expansion of the Universe.

These maps will be obtained by acquiring spectra with a relatively low signal-to-noise ratio and determining distances of the galaxies using a single emission line (and its photometric properties). Unfortunately, these two features combined may introduce significant systematics in the data analysis. 

The Hubble expansion of the Universe shifts the spectrum of galaxies towards lower frequencies (redshift), such that the rest-frame and observed spectra appear different. The redshift of each object is defined as $z =(\lambda_{\rm o}-\lambda_{\rm e})/\lambda_{\rm e}$, where $\lambda_{\rm o}$ and $\lambda_{\rm e}$ are the observed and emitted wavelengths of a specific line in the spectrum. When the rest-frame wavelength $\lambda_{\rm e}$ is known, the amount of shift of the spectrum is also determined; this can be used to infer the distance of emitting objects once a cosmology is assumed. This seems to be an easy procedure in principle, but it presents some caveats. Two lines emitted by separate objects A and B at different distances have different rest-frame wavelength $\lambda_e^A \neq \lambda_e^B$, but can have the same observed wavelength of $\lambda_o^A = \lambda_o^B$. If the two lines are misunderstood to have the same rest-frame wavelength (e.g. $\lambda_e^A$ is misunderstood to be $\lambda_e^B$), then the emission-line galaxy A will be associated with an incorrect redshift and thus placed at a wrong distance. This galaxy is called an interloper and it can present a source of systematic error, particularly when the measured spectra have a low signal-to-noise ratio. This presents even more significant problems when each galaxy redshift is computed with a single emission line (see e.g. \citealt{Pullen_2015, Gebhardt_2018, Addison_2018}).

The Roman Space Telescope will observe emission lines at 1.00--1.93 $\mu$m observed wavelength (\citealt{Spergel2015}; see update by \citealt{2019arXiv190205569A}). This includes H$\alpha$ emitters at $0.52<z<1.94$ and \oiii\ emitters at $1.00<z<2.85$. The \oiii\ feature is a doublet with the primary line at 500.7 nm and the secondary line at 495.9 nm, with a line ratio of 3:1 \citep{2000MNRAS.312..813S}. A doublet should be easier to distinguish from artifacts, noise, and other single emission lines. At the dispersion of the Roman grism ($\Delta\lambda_{\rm obs} = 2.17$ nm per 2 pixels) the \oiii\ doublet is resolved in most galaxies. However, low signal-to-noise spectra and small equivalent widths can make the detection of the second line very hard, and \oiii\ can become indistinguishable from a single line emission for some of the spectra.

The Roman Space Telescope will use the primary \oiii\ line to determine the distance of high redshift \oiii\ emission-line galaxies. This line is close to the H$\beta$ line (486.1 nm), which can become the source of interlopers. The difference between the true and the inferred radial position of an H$\beta$ galaxy misled as an \oiii\ emitter is
\begin{equation}
\Delta d \simeq 89 h^{-1}\, {\rm Mpc}\,\frac{1+z}{\sqrt{\Omega_{\Lambda}+\Omega_m(1+z)^3}}\,,
\label{eq:error_position}
\end{equation} 
where $z$ is the galaxy redshift, $\Omega_{\Lambda}$ and $\Omega_{m}$ are the cosmological constant and matter density parameters. Around redshift $z\sim 2$ this corresponds to roughly $90 h^{-1}\,$Mpc, which is a length scale close to, but less than, the Baryonic Acoustic Oscillation (BAO) scale. 

The BAO scale is a standard ruler that can be measured in the two-point correlation function of galaxies. The monopole of the correlation function presents a single peak at the BAO scale.  
This scale is proportional to $D_A(z)^2/H(z)$ and provides a degenerate measurement of the angular diameter distance $D_A(z)$ and the Hubble parameter $H(z)$ at the median redshift $z$ of the considered galaxy sample. The distance--redshift relation depends on the values of cosmological parameters. Therefore, measurements of the BAO scale (or peak) can be used to infer the expansion history of the universe and its cosmological contents. The presence of H$\beta$ interlopers is expected to produce an additional peak in the measured correlation function at $\sim 90 h^{-1}\,$Mpc, which blends with the BAO peak and thus results in a change in the best-fit shape and position. The presence of an additional shift due to interlopers can, therefore, lead to a biased estimation of cosmological parameters.

In this paper, we analyze the shift in BAO peak position in the ``\oiii'' sample of the Roman Space Telescope High Latitude Survey due to H$\beta$ interlopers. We study this effect at redshifts $1<z<2$ because we are limited by the maximum redshift reached by the currently available lightcone simulations that are populated with galaxy spectra. In this redshift range, the main Roman Space Telescope target is H$\alpha$ and its BAO analysis could be biased by line blending \citep{Martens:2018zvn}, while the \oiii-H$\beta$ confusion will be of minor impact in the cosmological analysis, but will still be present. Therefore, the main purpose of this paper is to present the phenomenon and the methodology to study it. Further analysis at higher redshift will be needed to understand the real impact of this line confusion in Roman Space Telescope data. 

For the analysis we use a lightcone $N$-body simulation --- described in Section~\ref{sec:sim} --- to generate a Roman-like \oiii\ mock galaxy catalog with H$\beta$ interlopers, as explained in Section~\ref{sec:cat}. In Section~\ref{sec:correlation} we present the 2-point correlation function measured in the catalogs, with and without interlopers. The formalism and results of the BAO fit --- a quantitative analysis of the BAO shift --- is shown in Section~\ref{sec:BAOfit}. Conclusions are drawn in Section~\ref{sec:conclusion}.

Even though this paper is tailored to Roman, the BAO shift due to \oiii-H$\beta$ confusion can be a common challenge in all future spectroscopic surveys (e.g. Euclid), and the methodology presented here can easily be generalized and applied to all of them.

\section{Simulation}
\label{sec:sim}
\begin{figure}
    \centering
	\includegraphics[width=0.9\columnwidth]{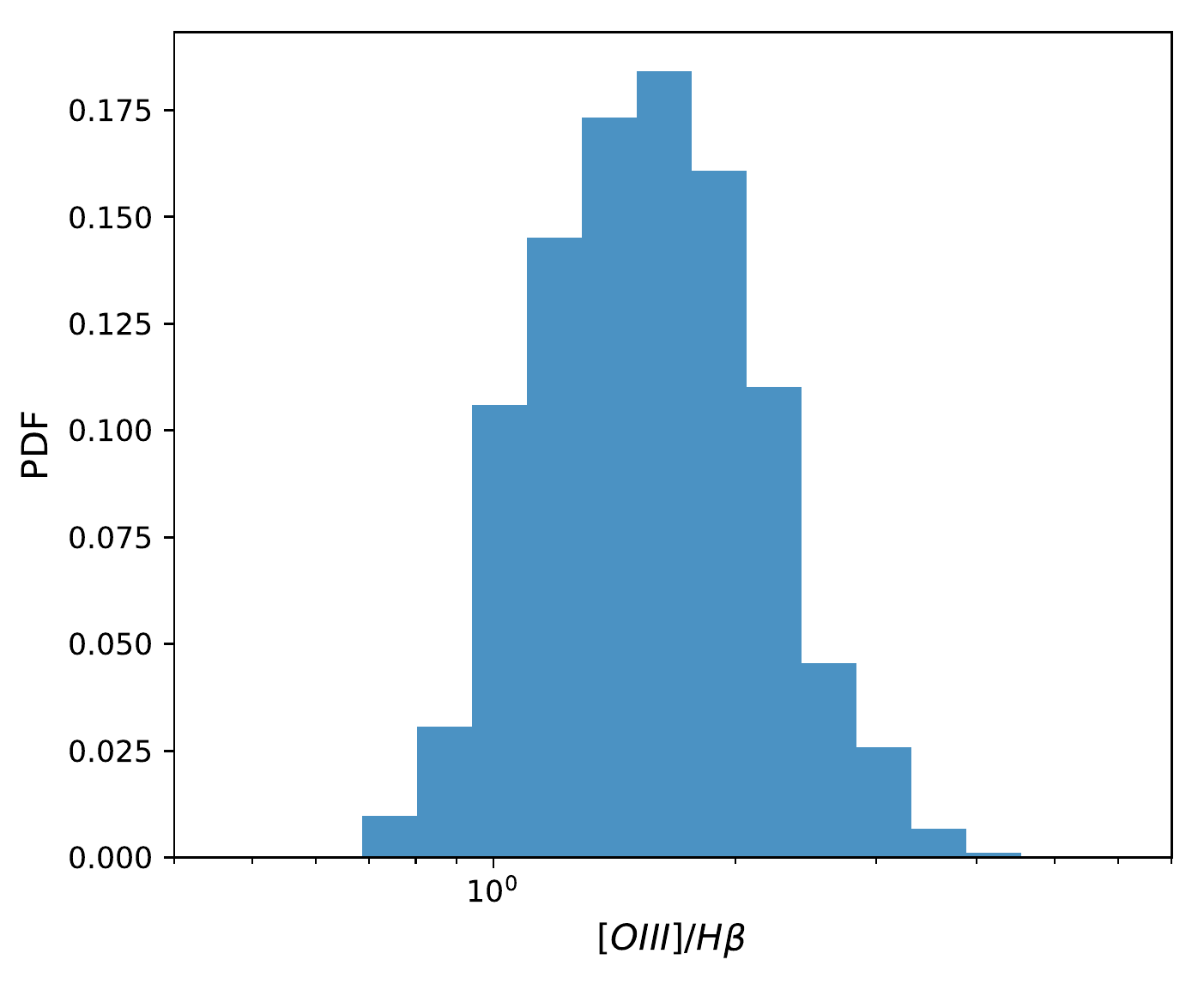}
    \caption{Probability distribution function of \oiii /H$\beta$ equivalent width ratio in the redshift range $1.3<z<1.9$ in the Roman-like mock catalog.}
    \label{fig:OIII_Hbeta_ratio}
\end{figure}
We use a lightcone simulation that covers 10,313 square degrees of the sky up to redshift $z=2.35$.  It has been generated from three $N$-body simulations with box-size $L=1.05, 2.6, 4.0 \,h^{-1}$Gpc and with number of dark matter particles $N=1400^3,2048^3,2048^3$. These three boxes are used to build the lightcone in redshift bins $0.0<z<0.32$, $0.32<z<0.84$ and $0.84<z<2.35$, respectively. The variation in resolution --- becoming lower with increasing redshift --- allows to decrease the computational cost of the simulations while achieving the minimum halo mass required for flux-limited surveys, that increases with redshift. The $N$-body simulations are run in a $\Lambda$CDM cosmology with $\Omega_{\rm m} = 0.286$, $\Omega_{\rm b} = 0.046$, $h=0.7$, $\sigma_8=0.82$, $n_s=0.96$, and three massless neutrino species ($N_{\rm eff}=3.046$). 

Galaxies are added to the lightcone using the ADDGALS algorithm \citep{DeRose2019, Wechsler2020}. ADDGALS is tuned using a Sub-Halo Abundance Matching (SHAM) scheme that reproduces the luminosity dependent clustering of the SDSS catalog with high precision. The algorithm assigns central galaxies to halos and subsequently adds the remaining galaxies to dark matter overdensities following a distribution of galaxy overdensities conditioned on absolute magnitude and redshift. The spectral energy distributions (SEDs) are assigned to each galaxy to match the SED-luminosity-density relationship measured in the SDSS data. The assignment is performed using a set of principle component coefficients as in \cite{2007AJ....133..734B}. We refer the reader to \cite{DeRose2019} for further details on the $N$-body simulations and to \cite{Wechsler2020} for more details on the galaxy model.

In what follows we will identify and use \oiii\ and H$\beta$ lines from the galaxy spectra. Fig.~\ref{fig:OIII_Hbeta_ratio} shows the probability distribution function (PDF) of the \oiii\ to H$\beta$ equivalent width ratios in the lightcone. The plot has been obtained by considering only galaxies that have both lines being detectable by the Roman Space Telescope (see Sec.~\ref{sec:cat} for more details on what this means). The measured ratio assumes values in the interval $[0.7\ ,5]$ and its distribution peaks around 1.5. The MOSDEF collaboration\footnote{\url{http://mosdef.astro.berkeley.edu}} observed emission-line galaxies in the same redshift range and measured larger \oiii$/$H$\beta$ ratios: their distribution peaks around 3 (see Fig.~\ref{fig:mosdef_EWratio}). This indicates that the lightcone could have more interlopers than the real data.  Future exploration will need to study this effect with samples that are directly tuned to this observable.

\section{Roman-like catalogs}
\label{sec:cat}
In this section we describe the construction of the Roman-like \oiii\ mock catalog, and how the H$\beta$ interlopers are added later on. The \oiii\ catalog is created by selecting specific galaxies from the lightcone. The selection criterion depends on two requirements. 

\begin{figure}
	\centering
	\includegraphics[width=0.8\columnwidth]{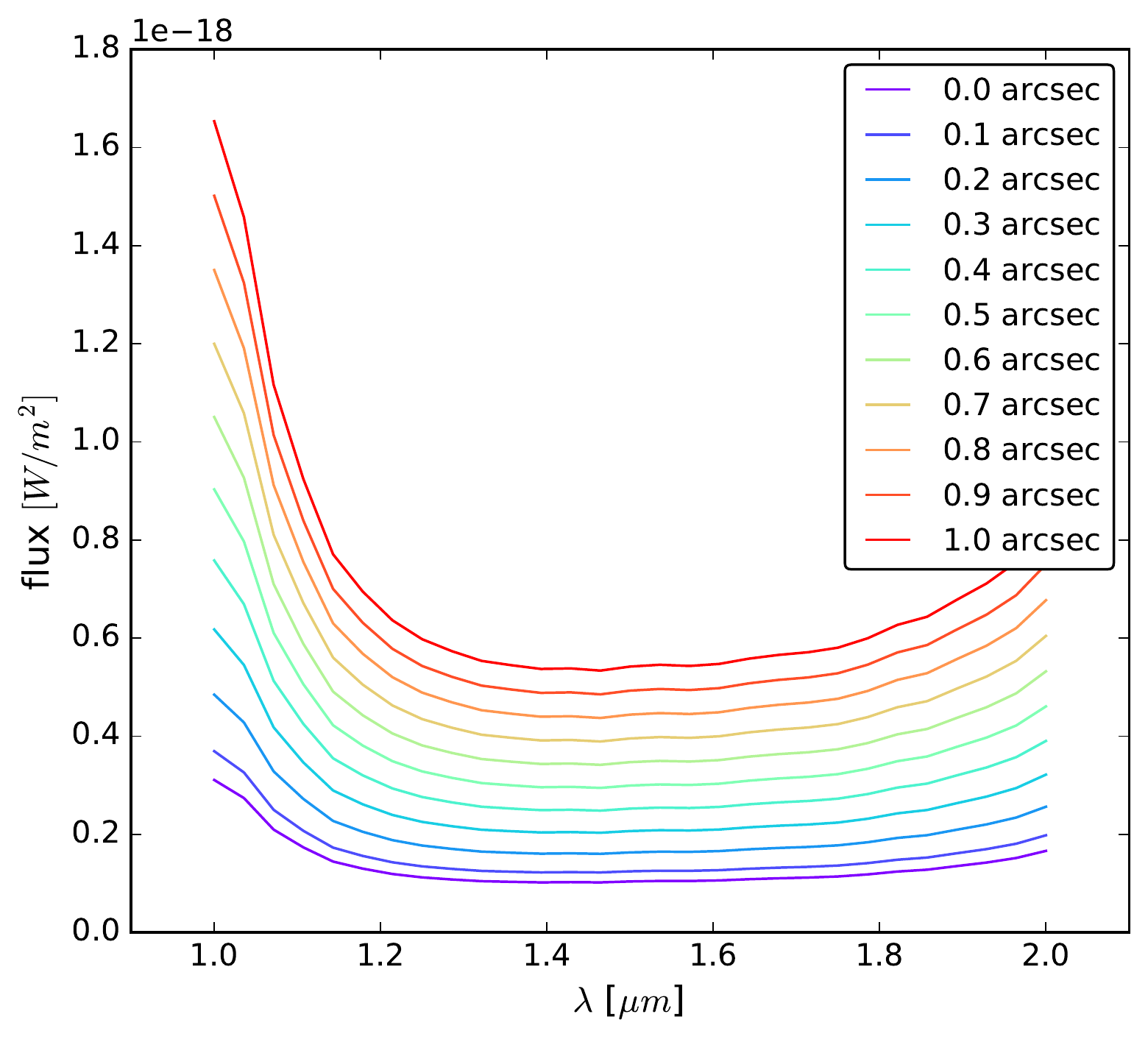}
    \caption{Roman flux detection limit as a function of observed wavelength and size of the emitting galaxy.
}
    \label{fig:detection_limit}
\end{figure}

The first requirement asks for the galaxy to be visible as an \oiii\ emitter by the Roman Space Telescope, i.e. 
it must have an \oiii\ flux above the Roman detection limit. We set the detection limit to $6.5 \sigma$ sensitivity at 6 exposure depth. 
This limit is displayed in Fig.~\ref{fig:detection_limit} as a function of the observed wavelength and the angular size of the emitted object. All objects with \oiii\ flux above detection limit are potentially \oiii\ emission-line galaxies to place in our catalog.

The second requirement asks for the catalog to reproduce the Roman forecast number density of \oiii\ emission-line galaxies in the redshift range of interest: $1.00<z<2.85$. However, the maximum redshift of the simulation used here is $z=2.35$, and above $z=1.8$ the number of galaxies drops drastically.
This is likely due to inaccuracies in the spectral properties of star-forming galaxies at these redshifts in the catalog.  Here we relax the second requirement by requiring the distribution of \oiii\ galaxies to reproduce the forecast in the redshift range $1.08 < z <1.8$.  The number density in the mock catalog is large than expected; thus we randomly sub-sample $11.4 \%$ of the the \oiii\ emission-line galaxies that formally meet the Roman detection limit. 

\begin{figure}
	\includegraphics[width=\columnwidth]{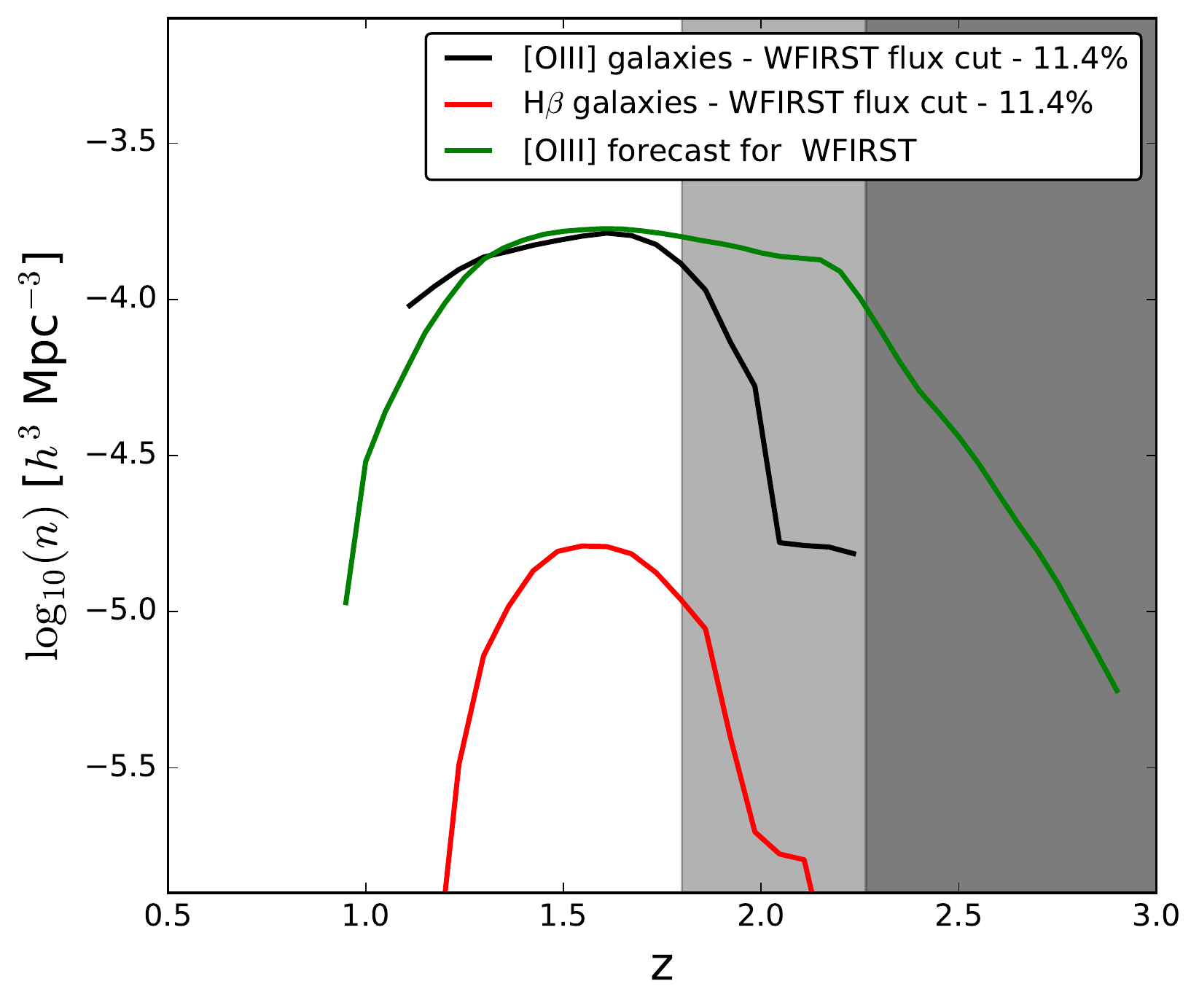}
    \caption{Redshift evolution of galaxy number density in the Roman catalog. The black line shows \oiii\ galaxies and the red line indicates the amount of H$\beta$ interlopers. The green line displays the predicted \oiii\ galaxies distribution in the Roman Space Telescope Survey. The light grey area indicates the redshift range where the number of \oiii\ emitters drops considerably---this interval has not been used to calibrate the \oiii\ mock catalog. The dark grey region shows the redshift range not reached by the lightcone simulation.}
    \label{fig:numDensity_OIII_Hbeta}
\end{figure}

Applying the Roman detection limit and a random sub-sampling applied over all redshifts, we have created 50 different Roman-like \oiii\ mock galaxy catalog. The galaxy distribution of one realization is shown in Fig.~\ref{fig:numDensity_OIII_Hbeta} (black line), together with the forecast distribution for \oiii\ from Roman Space Telescope (green line). 

We create a second type of mock catalogs, where H$\beta$ interlopers are added to the \oiii\ catalogs described above. To select the H$\beta$ emitters, we perform a galaxy selection in the lightcone simulation similar to what we have done for the \oiii\ emitters. First, we select galaxies with H$\beta$ flux above the Roman detection limit. Among these, some of them present both H$\beta$ and \oiii\ lines detectable, others have an \oiii\ flux too weak to be detected. The formers have two visible lines that can as a matter of principle be identified correctly, and are not of interest. The latter have a strong H$\beta$ line that will be mislead as \oiii\ and are interlopers. Thus, we select the H$\beta$ interlopers as galaxies with visible H$\beta$ and non-detectable \oiii\ lines. In order to be consistent with what has been done for the \oiii\ catalog, we randomly sub-sample to $11.4 \%$ the selected H$\beta$ galaxies. These interlopers are not placed in the mock catalog at their right position, but at a shifted position as if they were misled to be \oiii\ emitters. Their actual (wrong) redshift-position $\hat{z}$ in the mock is described by
\begin{equation}
1+\hat{z} = \frac{\lambda_e^{{\rm H}\beta}}{\lambda_e^{\rm [O III]}} \, (1+z),
\end{equation}
where $\lambda_e^{{\rm H}\beta}$ and $\lambda_e^{\rm [O III]}$ are the rest-frame wavelengths of H$\beta$ and \oiii, and $z$ is the true redshift of the galaxy. The distribution of H$\beta$ interlopers is shown in Fig.~\ref{fig:numDensity_OIII_Hbeta} (red line). Following this procedure and sub-sampling galaxies in 50 different ways, we have generated 50 different mocks with \oiii\ emission-line galaxies and H$\beta$ interlopers.

We have created the same type of catalogs in redshift space (RSD) by selecting \oiii\ and H$\beta$ emitters after the emission line spectra have been shifted to take into account also the redshift due to the peculiar velocity of the galaxy: $\lambda_o = (1+ z+ z_v) \lambda_e$, where $\lambda_o$ is the observed wavelength, $z$ is the true redshift and $z_v = v/c$ is the shift due to the peculiar velocity $v$ and $c$ is the speed of light.  In these catalogs, \oiii\ emission-line galaxies are placed at redshift $z+ z_v$ and H$\beta$ interlopers are placed at the misidentified redshift 
\begin{equation}
1+\hat{z}_{RSD} = \frac{\lambda_e^{{\rm H}\beta}}{\lambda_e^{\rm [O III]}} \, (1+z+ z_v).
\end{equation}

\section{Two-point correlation function}
\label{sec:correlation}

In this section, we present the two-point correlation function measured in the \oiii\ and \oiii+H$\beta$ catalogs. We perform the measurements in three redshift bins: $z=1.3-1.5$, $1.5-1.7$, and $1.7-1.9$. The percentage of H$\beta$ interlopers over the total number of galaxies in these bins are 7.7\%, 9.04\%, and 7.96\%, respectively.

We calculate the galaxy correlation function using the Landy--Szalay estimator
\begin{equation}
\hat{\xi}(r,\mu) = \frac{DD(r,\mu)-2DR(r,\mu)/f_R+RR(r,\mu)/f_R^2}{RR(r,\mu)/f_R^2}\, ,
\end{equation}
where $DD$, $DR$ and $RR$ are the number of galaxy--galaxy, galaxy--random and random--random pairs at separation $r$, and $\theta=\arccos(\mu)$ is the angle between the separation vector and the line of sight. The random catalog $R$ is generated from a set of random points in a volume identified by the redshift selection in the galaxy catalogs. The redshift distribution of the randoms reproduces the distribution of the galaxies and has been tuned so that the ratio $f_R$ between the number of galaxies and the number of random points is $f_R=10$. The pair counting is performed using the large-scale structure toolkit \textsc{nbodykit} (\citealt{Hand:2017pqn}).

We project the two-point correlation function $\hat{\xi}$ in a basis of Legendre polynomials $L_l(\mu)$ of order $l$ via
\begin{equation}
\hat{\xi}_l(r) = \frac{2l+1}{2}\int_{-1}^1d\mu \,\hat{\xi}(r,\mu) \, L_l(\mu)\, ,
\end{equation}
where $l=\{0,2,...\}$ correspond to the Legendre polynomials $L_l=\{1,(3\mu^2- 1)/2,...\}$ and they identify the monopole, quadrupole, etc. of the correlation function.

In Fourier space, the multipoles of the power spectrum are
\begin{equation}
\label{eq:P_l}
P_l(k) = \frac{2l+1}{2}\int_{-1}^1d\mu \,P(k,\mu) \, L_l(\mu)\, ,
\end{equation}
and they are related to the multipoles of the two-point correlation function via 
\begin{equation}
\label{eq:xi_l}
\xi_l(r) = i^l \int_0^\infty \frac{k^2dk}{2\pi^2}P_l(k)j_l(k r)\,,
\end{equation}
with $j_l(k r)$ being the spherical Bessel function of order $l$. 

\begin{figure}
	\includegraphics[width=\columnwidth]{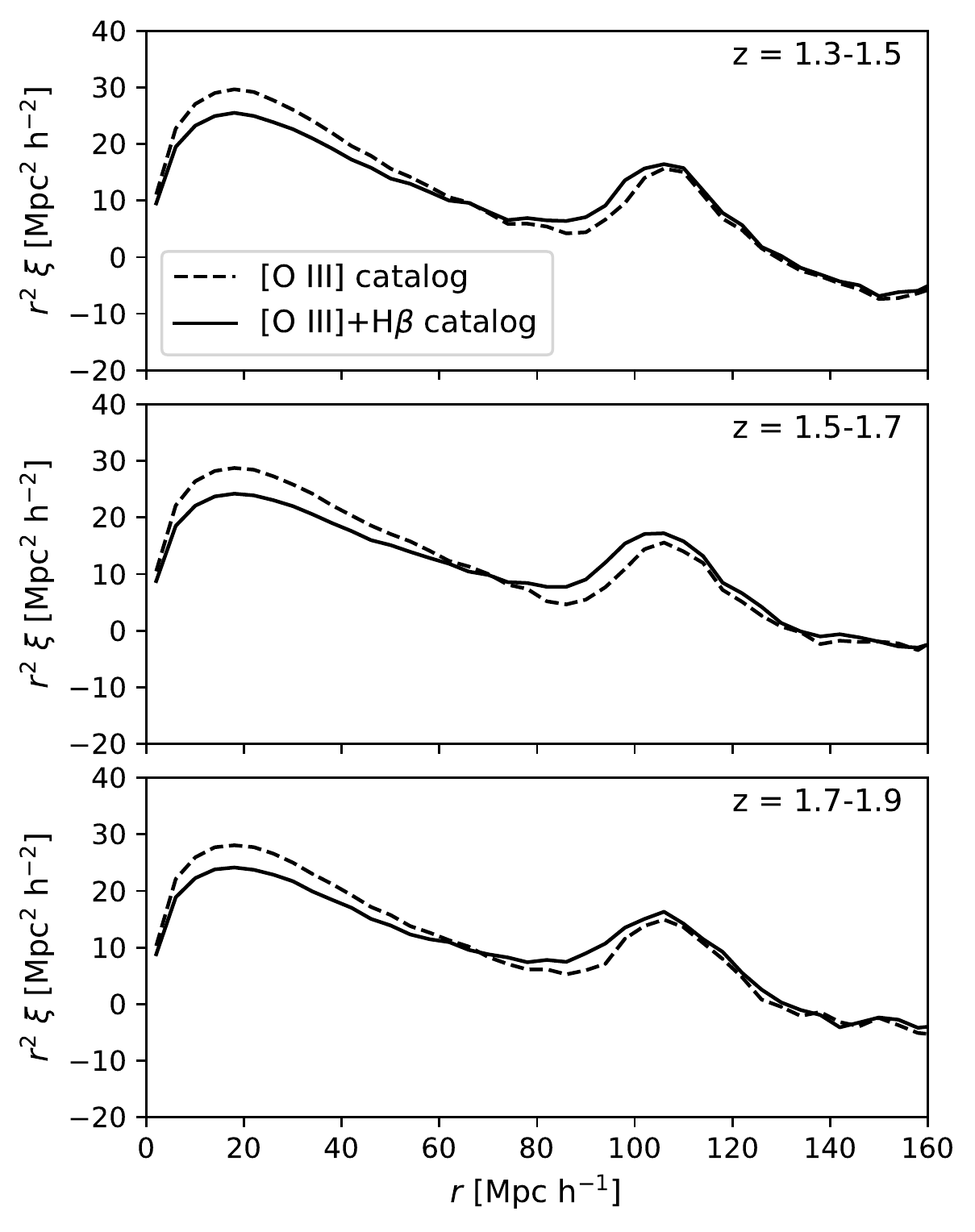}
    \caption{Mean of the two-point correlation function in real space measured in $50$ realizations. Dashed lines show the correlation of [OIII] galaxies, while solid lines indicate the correlation in the full catalog, including both [OIII] and H$\beta$ galaxies.
}
    \label{fig:xi_RealSpace}
\end{figure}

Figure~\ref{fig:xi_RealSpace} shows the mean galaxy correlation function computed from $50$ mocks in real space. Dashed lines indicate the measurements in the \oiii\ catalogs, 
while solid lines show the measurements in the \oiii+H$\beta$ catalogs.  The presence of interlopers modifies the correlation function by suppressing it on small separations and enhancing it around the BAO scale. The BAO peak appears broadened and shifted towards smaller separations. This behavior is understood by considering a small volume containing both \oiii\ and H$\beta$ galaxies, which will be correlated and sourcing the correlation on small separation. In our analysis, the H$\beta$ galaxies have been moved away from the considered volume by $\sim 90h^{-1}$Mpc. This means that there is a lack of galaxy pairs on small separation and an increase in the number of galaxy pairs with separations close to the BAO scale.

\begin{figure}
	\includegraphics[width=\columnwidth]{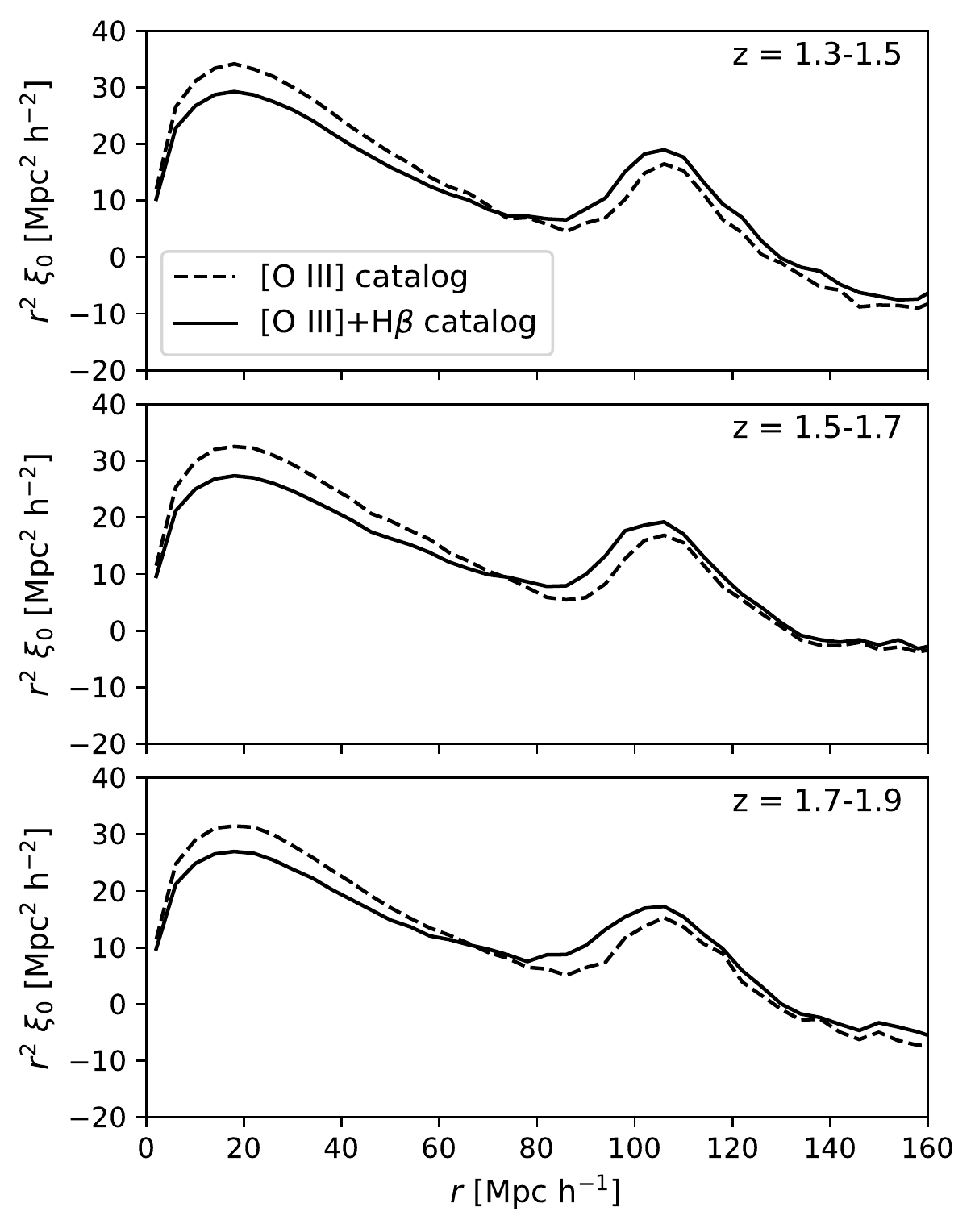}
    \caption{Mean of the two-point correlation function in redshift space measured in $50$ realizations. Dashed and solid lines are as in Fig.~\ref{fig:xi_RealSpace}. 
}
    \label{fig:xi_RSD}
\end{figure}

Fig.~\ref{fig:xi_RSD} shows the analogous measurements in redshift space. The BAO peak is broadened and shifted towards small scales in this case as well. 

\section{BAO fitting}
\label{sec:BAOfit} 
In this section, we quantify the BAO shift due to H$\beta$ interlopers. Firstly, we build a model for the monopole of the correlation function in the true cosmology of the simulation, allowing for a free parameter to describe the BAO shift. Secondly, we fit this model to the measured \oiii+H$\beta$ monopole to determine the BAO shift induced by the interlopers. 

\subsection{Model}
The BAO analysis of the monopole of the correlation function allows to measure the spherically averaged distance 
\begin{equation}
D_v(z) = \left[(1+z)^2D_A^2(z)\frac{cz}{H(z)} \right]^{1/3}.
\end{equation}
The interloper-induced shift of the BAO scale relative to the true value can be described by the isotropic dilation parameter
\begin{equation}
\alpha = \frac{D_{v,i}(z)/r_{s,i}}{D_v(z)/r_s}= \left[ \frac{D_{A,i}^2(z)}{D_A^2(z)} \frac{H(z)}{H_i(z)} \right]^{1/3}\frac{r_s}{r_{s,i}}\,,
\label{eq:alpha}
\end{equation}
where $r_s$ is the BAO scale (sound horizon) and the subscript $i$ indicates the quantities in the presence of H$\beta$ interlopers. Quantities without the subscript $i$ correspond to the true cosmology used to compute all the measurements, also when interlopers are present. 

The parameter $\alpha$ is often used to quantify the shift between the BAO scale in the fiducial and the true (inferred from the data) cosmology. Analogously to this case, we define a fitting model for the correlation function given by
\begin{equation}
\xi_{\rm m}(r) = B(r)\xi(r\alpha)+A(r)\,,
\end{equation}
where $\xi$ is the template for the nonlinear galaxy correlation function based on linear theory and on the true cosmology, $\alpha$ is the isotropic dilation parameter in Eq.~(\ref{eq:alpha}) used to adjust the location of the BAO peak and, $A(r)$ and $B(r)$ are functions involving nuisance parameter used to marginalize over broadband effects of the correlation function, such as scale-dependent bias and redshift-space distortion effects. Using $A(r)$ and $B(r)$ should minimize the contribution of broadband effects to the BAO scale and should allow the fitting model to make robust and broadband-independent predictions on $\alpha$. In our analysis we use
\begin{equation}
A(r) = \frac{a_1}{r^2}+\frac{a_2}{r}+a^3 ~~{\rm and}~~
B(r) = B.
\end{equation}
We model the template for the nonlinear galaxy correlation function from its Fourier transform, the power spectrum $P(k,\mu)$, given by
\begin{equation}
\label{eq:P_rsd}
P(k,\mu) = b^2 \,(1+\beta \mu^2)^2\, F(k,\mu,\Sigma_s)\,P_{w}(k,\mu)\,,
\end{equation}
where $b$ is the linear galaxy bias, $(1+\beta \mu^2)^2$ is the Kaiser term describing the large-scale redshift-space distortions with $\beta = f/b$ and $f=\Omega_m^{0.55}$ being the linear growth rate. The function 
\begin{equation}
F(k,\mu,\Sigma_s) = \frac{1}{(1+k^2\mu^2\Sigma_s^2)^2}
\end{equation}
is the streaming model describing the Finger of God (FoG) effect, with $\Sigma_s$ being the streaming scale of order $\sim 3-4h^{-1}$Mpc. The wiggled power spectrum $P_{w}(k,\mu)$ is defined as
\begin{equation}
P_{\rm w}(k,\mu) = [P_{\rm lin}(k)-P_{\rm nw}(k)]
\exp \left(-k^2 \Sigma_{\rm nl}^2/2\right)+P_{\rm nw}(k)\,,
\end{equation}
where $P_{\rm lin}$ is the linear matter power spectrum, $P_{\rm nw}$ is the no-wiggle counterpart computed as in \cite{Vlah2015}, and $\Sigma_{\rm nl}^2 = \Sigma_{0}^2(1 + (2f+f^2)\mu^2) $ (\citealt{Cohn:2015ljb}) with 
\begin{equation}
\Sigma^2_{0} = \int \frac{dk}{ 3 \pi^2} \left[ 1- j_0(k r_s) \right] P_{\rm lin}(k)  
\end{equation}
(\citealt{Vlah2015})
describes the BAO damping due to nonlinear structure growth. The power spectrum in Eq.~(\ref{eq:P_rsd}) can be decomposed in multipoles using Eq.~(\ref{eq:P_l}); they are related to the multipoles of the correlation function via Eq.~(\ref{eq:xi_l}).

\subsection{Covariance matrix}
To compute the BAO fit, we need to estimate the covariance matrix of the two-point correlation function. Ideally, it should be computed from a set of mock catalogs generated from independent realizations of the Universe via
\begin{equation}
\label{eq:cov_mock}
C_{ij}[\xi(r_i)\,\xi(r_j)] = \frac{1}{N-1}\sum_{n=1}^N \,[\xi_n(r_i)-\bar{\xi}(r_i)][\xi_n(r_j)-\bar{\xi}(r_j)].
\end{equation} 
However, we have only one realization from which we randomly subsampled the galaxies in $50$ different ways to generate $50$ different mock galaxy catalogs. We could use them to compute the covariance matrix as in Eq.~(\ref{eq:cov_mock}), but this procedure is not accurate since the mocks are not generated from independent realizations of the dark matter field. Therefore, we compute the covariance matrix analytically. For comparison, the results using the covariance matrix of the 50 mocks are presented and discussed in Appendix~\ref{sec:cov_50mocks}. 

The simplest analytical model for the covariance matrix is the Gaussian model, motivated by the fact that matter fluctuations are Gaussian in the initial conditions. In this case, the covariance between two multiple moments $l$ and $l'$ of the correlation function is (\citealt{Xu:2012hg})
\begin{eqnarray}
\label{eq:cov_theory}
C_{ij}(\xi_l(r_i)\xi_{l'}(r_j)) &=& \frac{2(2l+1)(2l'+1)}{V}\,i^{l+l'}\\\nonumber
&& \times\int\frac{k^2dk}{2\pi^2}j_l(kr_i)j_{l'}(kr_j)P_{ll'}^2(k)\,,
\end{eqnarray}
where $V$ is the volume considered, $j_{l}$ is the spherical Bessel function of order $l$ and
\begin{equation}
P_{ll'}^2(k) = \frac{1}{2}\int_{-1}^1\left[ P(k,\mu)+\frac{1}{\bar{n}} \right]^2 L_l(\mu) L_{l'}(\mu) d\mu
\end{equation}
contains the power spectrum $P(k,\mu)$ in Eq.~(\ref{eq:P_rsd}) and the Poisson shot-noise $1/\bar{n}$. We take into account the redshift dependence of $\bar{n}$ and assume it has no angular dependence by considering the quantity
\begin{equation}
I^2(k) = \int\frac{dV}{P_{ll'}^2(k)}
\end{equation}
with 
\begin{equation}
dV = \frac{c}{H_0}\frac{r^2(z)}{\sqrt{\Omega_m(1+z)^3+\Omega_{\Lambda}}}\,dz\,d\Omega \,,
\end{equation}
and by replacing $P_{ll'}^2(k)/V$ with $[I^2(k)]^{-1}$ in Eq.~(\ref{eq:cov_theory}). 

The correlation function measured in the mocks is binned and we must account for it in the modeling of the Gaussian covariance matrix. If the correlation functions are measured in bins with lower bounds $r_1$ and upper bound $r_2$, then the binned Gaussian covariance matrix is
\begin{eqnarray}\nonumber
C_{ij}(\xi_l(r_i)\xi_{l'}(r_j)) \!\!&=&\!\! 2(2l+1)(2l'+1)\,i^{l+l'}\frac{3}{r_{i2}^3-r_{i1}^3}\frac{3}{r_{j2}^3-r_{j1}^3}\\\nonumber
&& \times \int_{r_{i1}}^{r_{i2}}r^2 dr \frac{d\Omega}{4\pi}\int_{r_{j1}}^{r_{j2}}r^2 dr \frac{d\Omega}{4\pi}\\
&& \times\int\frac{k^2dk}{2\pi^2}j_l(kr_i)j_{l'}(kr_j)[I^2(k)]^{-1}.
\end{eqnarray}
In this work we use only the monopole to calculate the BAO fit. In this case $l=l'=0$ and the above binned covariance matrix becomes
\begin{equation}
C_{ij}(\xi_l(r_i)\xi_{l'}(r_j)) = 2 \int\frac{k^2dk}{2\pi^2}\Delta_l(kr_i)\Delta_{l'}(kr_j)\,[I^2(k)]^{-1}\,,
\end{equation}
where 
\begin{equation}
\Delta(kr) = \frac{3}{r_{2}^3-r_{1}^3}\left[ \frac{r_2^2\, j_1(kr_2)-r_1^2\, j_1(kr_1)}{k}\right] 
\end{equation}
and $j_1(kr)$ is the spherical Bessel function of order $1$.

There have been studies aiming to modify the simple Gaussian model to create more precise covariance matrices, see e.g. \cite{Xu:2012hg}. These approaches require the fit of additional parameters to the covariance matrix from many $N$-body simulations, or in our case many lightcones with emission-line galaxies. The limited number of lightcones available do not allow us to use these models.

\subsection{Results}

\begin{table}
\begin{center}
\begin{tabular}{| l | c | c | c |}
\hline
Case & $1.3 < z < 1.5$ & $1.5 < z < 1.7$ & $1.7 < z < 1.9$ \\
\hline
\hline
Real: $f_{H\beta} = 0$ & $0.995 \pm 0.012$ & $1.000 \pm 0.010$ & $1.001 \pm 0.012$\\
\hline
Real: $f_{H\beta} = 0.5$ & $1.002 \pm 0.010$ & $1.010 \pm 0.010$ & $1.004 \pm 0.016$\\
\hline
Real: $f_{H\beta} = 1$ & $1.007 \pm 0.011$ & $1.018 \pm 0.011$ & $1.012 \pm 0.016$\\
\hline
\hline
RSD: $f_{H\beta} = 0$ &  $0.996 \pm 0.011$ & $1.002 \pm 0.013$ & $0.998 \pm 0.014$\\
\hline
RSD: $f_{H\beta} = 0.5$ & $1.003 \pm 0.013$ & $1.010 \pm 0.011$ & $1.008 \pm 0.015$\\
\hline
RSD: $f_{H\beta} = 1$ & $1.008 \pm 0.010$ & $1.023 \pm 0.011$ & $1.021 \pm 0.015$\\
\hline
\hline
\end{tabular}
\caption{\label{tab:BAO_fit}Values of the isotropic dilation parameter $\alpha$ from the BAO fit in real and redshift space with different fractions $f_{H\beta}$ of interlopers. $f_{H\beta}=0$ corresponds to no interlopers, $f_{H\beta}=1$ indicates the inclusion of all the interlopers present in the mock catalog, and $f_{H\beta}=0.5$ is the case where half of the interlopers are considered.}
\end{center}
\end{table}

We perform a null test to validate our pipeline.  Here we perform the BAO analysis on the \oiii\ mock galaxy catalogs, where all galaxies are at the right redshift. In this case, we expect the parameter $\alpha$ to be close to $1$, as this indicates no shift of the BAO peak compared to the theoretical prediction in the cosmology of the simulation. The values for the best fit of $\alpha$ are shown in Table \ref{tab:BAO_fit}. There, $f_{H\beta}$ indicates the fraction of H$\beta$ interlopers. $f_{H\beta}=0$ corresponds to no interlopers (\oiii\ catalog), and $f_{H\beta}=1$ indicates the inclusion of all interlopers present in the mock catalog \oiii+H$\beta$. The values of $\alpha$ in the case $f_{H\beta}=0$ are $|\alpha-1|<0.5\%$ and are consistent with $1$, both in real (1st row) and redshift (4th row) space. 

The results for the BAO analysis applied on the \oiii+H$\beta$ catalog are shown in the same table and correspond to $f_{H\beta}=1$. In this case, $\alpha$ presents a systematic shift towards higher values. The shift is more pronounced in the redshift bin $1.5<z<1.7$, where the percentage of interlopers over the total number of galaxies is the highest. 

In order to understand how the shift depends on the number of interlopers in the three redshift bins, we performed the BAO analysis in an intermediate case, where only half of the interlopers are considered. We randomly subsample them and recompute the correlation function. The best fits for $\alpha$ are shown in Table \ref{tab:BAO_fit} as the case $f_{H\beta}=0.5$. Again, the BAO peak appears to be shifted, but the shift is smaller than in the case $f_{H\beta}=1$, where all interlopers were included. As expected, the BAO shift increases with the number of interlopers present in the catalog. 

\begin{figure}	\includegraphics[width=\columnwidth]{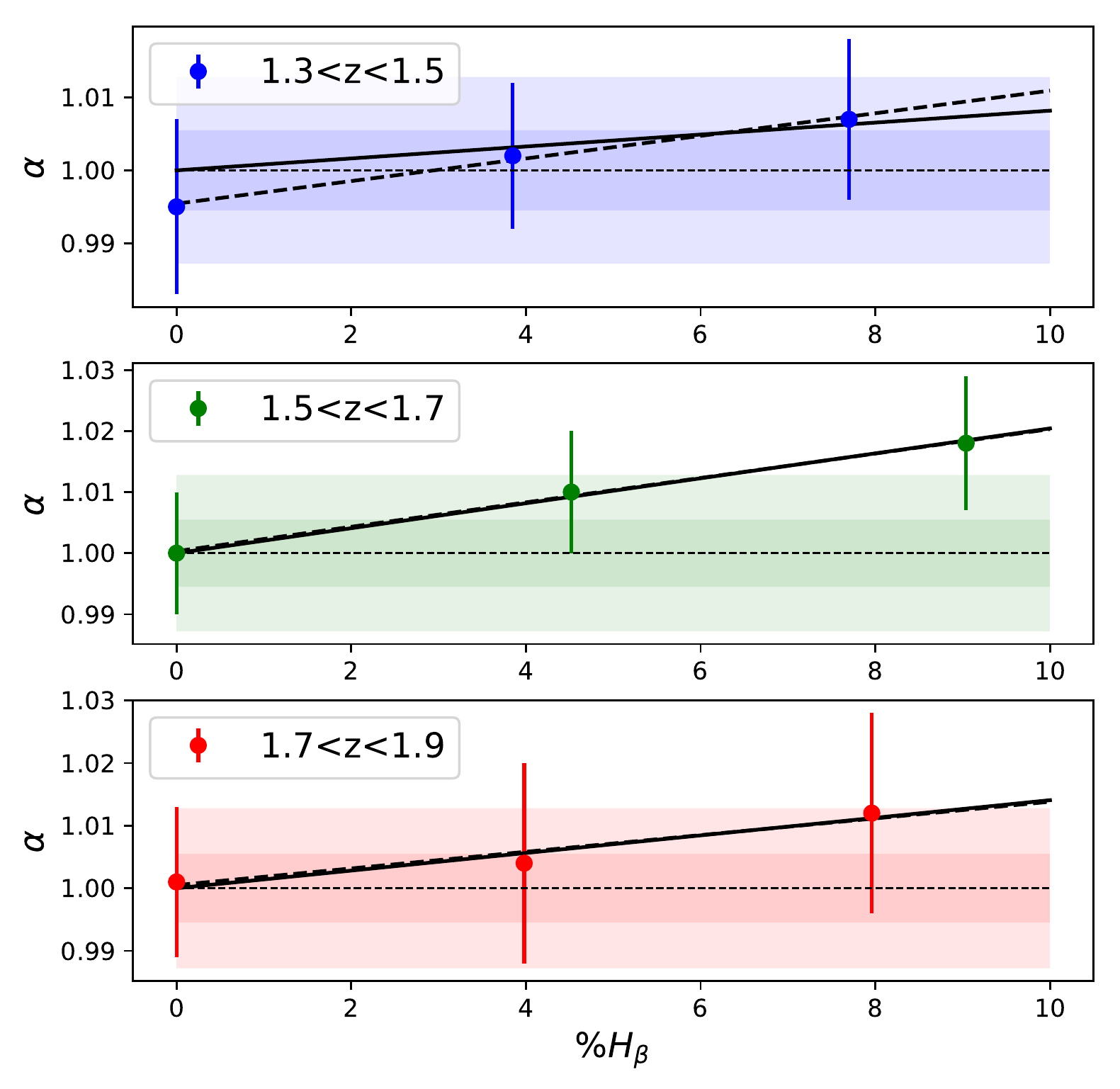}
    \caption{Results for the isotropic parameter $\alpha$ from the BAO fit in real space. $\% H_{\beta}$ indicates the percentage of interlopers over the total number of galaxies. $\% H_{\beta} = 0$ correspond to no interlopers and the $\% H_{\beta} >0$ values correspond to half and the complete number of interlopers in the mocks, for each redshift bin. Black solid lines show the linear fit $y = 1 + {\rm b}\,x $ (see also Table \ref{tab:b_fit}) while back dashed lines display the linear fit $y = a + {\rm b}\,x $, with $y=\alpha$ and $x=\% H_{\beta}$.}
    \label{fig:alpha_realSpace}
\end{figure}

\begin{figure}
\includegraphics[width=\columnwidth]{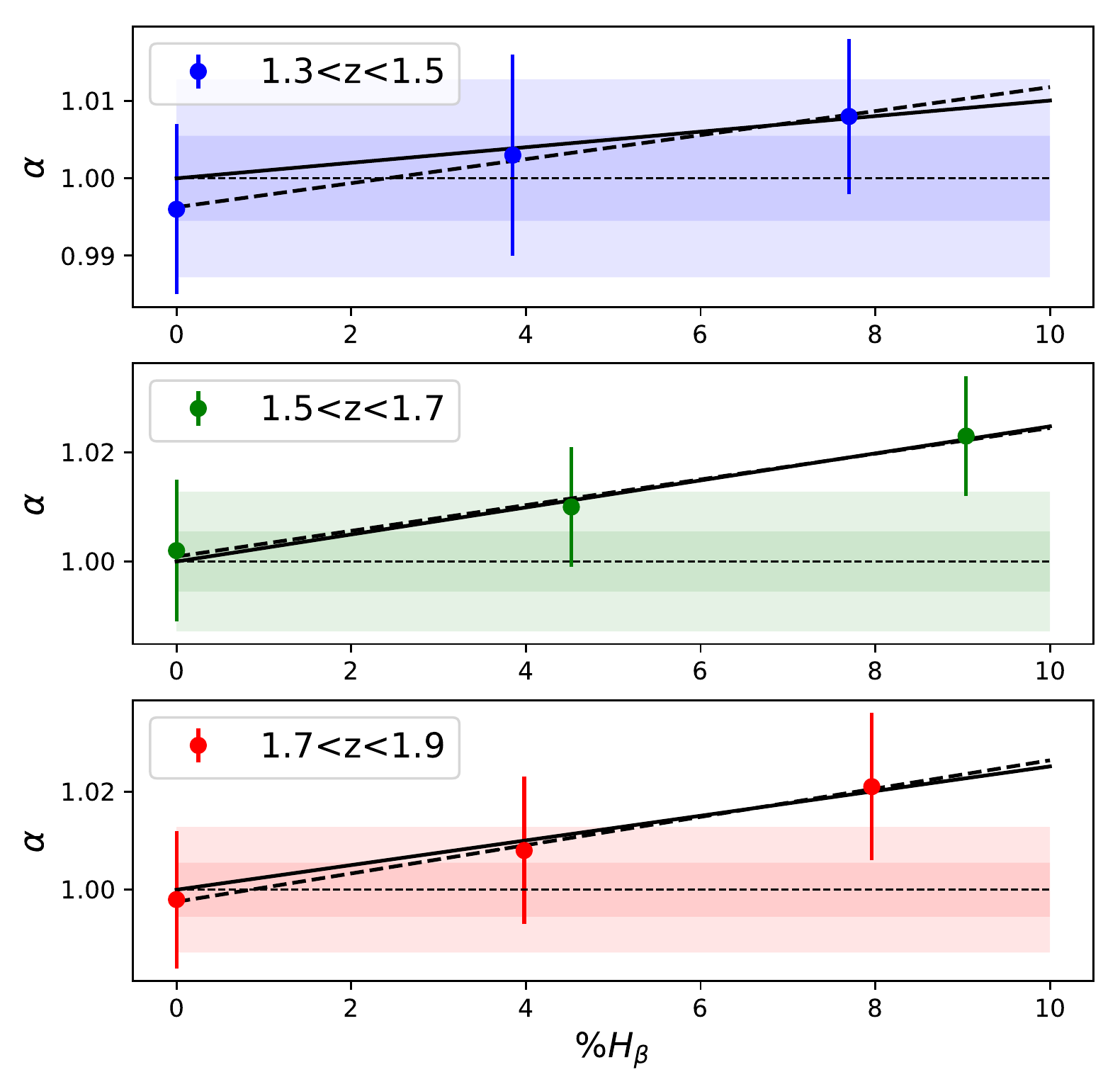}
    \caption{Results for the isotropic parameter $\alpha$ from the BAO fit in redshift space. The color code is the same as in Figure \ref{fig:alpha_realSpace}.}
    \label{fig:alpha_RSD}
\end{figure}

Figs. \ref{fig:alpha_realSpace} and \ref{fig:alpha_RSD} present a visualization of the shift as a function of the percentage of interlopers $\%H\beta$ over the total number of galaxies, in real and redshift space. Each panel shows different redshift bins. From left to right, each point correspond to $f_{H\beta}=0, 0.5, 1$, respectively. The dark shaded areas indicate the expected statistical error for $\alpha$ in the BAO analysis with \oiii\ emission-line galaxies in the Roman  Space  Telescope, $\sigma_\alpha = 0.55\%$  (\citealt{Spergel2015}; this number was updated using the number densities in preparation for the Reference Survey as described in \citealt{Eifler2020}). The light shaded areas show a more conservative estimation of the error: the one reported in the Roman Space Telescope Science Requirement Document ($\sigma_\alpha = 1.28\%$). 

To roughly estimate the BAO shift at different $\%H\beta$, we interpolate the three values of $\alpha$ in each redshift bin with a linear function. Since $\alpha$ should be equal to $1$ when $\%H\beta=0$, the linear function is forced to pass through the point (0,1), and the only free parameter is the slope. 
The best fits are displayed in Table \ref{tab:b_fit} and as black solid lines in Figs. \ref{fig:alpha_realSpace} and \ref{fig:alpha_RSD}.
The slope appears to be maximum in the redshift bin $1.5<z<1.7$ in real space, and at both $1.5<z<1.7$ and $1.7<z<1.9$ in redshift space. We already noticed that the parameter $\alpha$ is larger in the redshift range $1.5<z<1.7$. The linear fit shows this is not due to the larger amount of interlopers, but it is caused by the particular redshift range considered. Indeed, the error on the interloper positions depends on $z$ via Eq.~\ref{eq:error_position}, and the BAO shift will consequently depend on the redshift considered.  

The linear function gives also an estimation of the largest amount of interlopers that will produce a BAO shift smaller than a desired systematic errors on $\alpha$. Usually, the systematic errors $\sigma_{\rm sys}$ are required to be smaller than some percentage of the statistical error $\sigma_{\rm stat}$. If we consider $\sigma_{\rm sys} < \sigma_{\rm stat}/\sqrt{10}$ with $\sigma_{\rm stat} = 0.55\%$, then the maximum tolerated percentages of interlopers are $\sim 2.1\%,0.8\%,1.2\%$ in real space and $\sim 1.7\%,0.7\%,0.7\%$ in redshift space at $z\sim 1.4, 1.6,1.8$, respectively.

We perform a second linear fit to evaluate the theoretical uncertainties --- non-linearities, bias, and inaccuracies in the fitting function --- in our BAO analysis. This fit is performed by letting the intercept free to vary (black dashed lines in Figs. \ref{fig:alpha_realSpace} and \ref{fig:alpha_RSD}), since the theoretical uncertainties would make $\alpha\neq1$ at $f_{H\beta}=0$. The best fits for the intercept are all consistent with 1 within their error bars, but their value is more than $0.2\%$ far from 1 in the redshift bin $1.3 < z < 1.5$, both in real and redshift space. In these two cases the value of the intercept is smaller than 1, which indicates that our model could be underestimating the values of $\alpha$ in this redshift bin. The value of the slope appears to be slightly larger than the one obtained when fixing the intercept to 1 (Table \ref{tab:b_fit}). This indicates that our estimation of the maximum tolerated percentages of interlopers in $1.3 < z < 1.5$ could be underestimated.

\begin{table}
\begin{center}
\begin{tabular}{| l | c |}
\hline
Space \& redshift & b slope \\
\hline
\hline
Real: $1.3 < z < 1.5$ & $(8.19 \pm 8.85) \cdot 10^{-4}$\\
\hline
Real: $1.5 < z < 1.7$ & $(2.04 \pm 0.75) \cdot 10^{-3}$\\
\hline
Real: $1.7 < z < 1.9$ & $(1.41 \pm 1.27) \cdot 10^{-3}$\\
\hline
\hline
RSD: $1.3 < z < 1.5$ & $(1.00\pm 0.86) \cdot 10^{-3}$\\
\hline
RSD: $1.5 < z < 1.7$ & $(2.48 \pm 0.77) \cdot 10^{-3}$\\
\hline
RSD: $1.7 < z < 1.9$ & $(2.51 \pm 1.19) \cdot 10^{-3}$\\
\hline
\hline
\end{tabular}
\caption{\label{tab:b_fit} Results for the linear fit $y = 1 + {\rm b}\,x $, where $y$ is the isotropic parameter $\alpha$ and $x$ is the percentage of interlopers $\%H\beta$, in real and redshift space.}
\end{center}
\end{table}


\section{Conclusions}
\label{sec:conclusion}

In this paper we studied the impact that H$\beta$-\oiii\ line confusion can have on the BAO peak of \oiii\ emitters. We restrict the analysis to the monopole of the \oiii\ galaxy correlation function. Confusing an H$\beta$ emitter as an \oiii\ emitter introduces an error in the estimation of the galaxy redshift, which makes the galaxy appear to be closer to us than it actually is. This error is due to a modification of the \oiii\ galaxy correlation function's shape. In particular, the BAO peak in this correlation function becomes broader and shifted towards smaller scales. This happens because H$\beta$ interlopers add correlation power primarily on scales equal to the length of the displacement between their true and inferred position, which is $\sim 90$ Mpc/$h$.

To study this phenomenon, we have generated Roman-like mock catalogs containing \oiii\ galaxies and \oiii+H$\beta$ interloper galaxies. Then, we performed a BAO fit on the monopole measured in these mocks to quantify the impact of line confusion on future BAO analysis. 

The \oiii\ to H$\beta$ equivalent width ratio is not exactly known at the considered redshifts --- it is estimated to be around 4-5\% from the data obtained by the MOSDEF collaboration (see Fig.~\ref{fig:mosdef_interlopers}). Therefore, we cannot give an estimation of the expected BAO shift due to H$\beta$ interlopers. 
Instead, we estimate the shift as a function of fraction of H$\beta$ interlopers in the considered emission-line galaxy sample. If the Roman statistical error on $\alpha$ is $\sigma_{\rm stat} = 0.55\%$ and the systematic error introduced by interlopers must be $\sigma_{\rm sys} < \sigma_{\rm stat}/\sqrt{10}$, then the maximum tolerated percentages of interlopers are $\sim2.1\%,0.8\%,1.2\%$ in real space and $\sim 1.7\%,0.7\%,0.7\%$ in redshift space at $z\sim 1.4, 1.6,1.8$, respectively.  This phenomenon should be more problematic at higher redshift, $2<z<3$, where the \oiii\ line is the main target. However, in this redshift range the \oiii\ to H$\beta$ equivalent  width ratio is expected to be larger (see Fig.~\ref{fig:mosdef_EWratio}) and the interloper fraction is expected to be smaller (see Fig.~\ref{fig:mosdef_interlopers}).

Here we presented the methodology to study the impact of H$\beta$ interlopers on the monopole analysis, and delineate how to determine the level of interlopers that can be tolerated without introducing a systematic error. The effect of interlopers on the correlation function is anisotropic, thus the inclusion of the quadrupole in the BAO analysis is expected to help disentangling the interloper effect from the impact of a wrong fiducial cosmology, making H$\beta$ interlopers a less severe source of systematic error. Moreover, having a theory model to describe how H$\beta$ interlopers affect the quadrupole will allow to determine the amount of interlopers in a given galaxy sample. We will combine the analysis on monopole and quadrupole and present a model to describe H$\beta$ interlopers in a following paper, where we will also investigate the phenomenon at higher redshifts. 

\section*{Acknowledgements}
E.M. thanks Naveen Reddy for providing the \oiii\ and H$\beta$ equivalent width of the galaxies in the MOSDEF survey. She thanks Zachary Slepian and the Roman Science Investigation Team for useful discussions. She was supported by the NASA grant 15-WFIRST15-0008 during part of this project.
S.H. thanks NASA for their support in grant number: NASA grant 15-WFIRST15-0008 and
NASA Research Opportunities in Space and Earth Sciences grant 12-EUCLID12-0004. Both E.M. and S.H. thank Simons Foundation for supporting their work.  C.H. is supported by NASA grant 15-WFIRST15-0008 and the Simons Foundation. This research used resources of the National Energy Research Scientific Computing Center (NERSC), a U.S. Department of Energy Office of Science User Facility operated under Contract No. DE-AC02- 05CH11231. 




\bibliographystyle{mnras}
\bibliography{biblio} 





\appendix
\section{Covariance matrix from the 50 realizations}
\label{sec:cov_50mocks}

\begin{figure}	\includegraphics[width=\columnwidth]{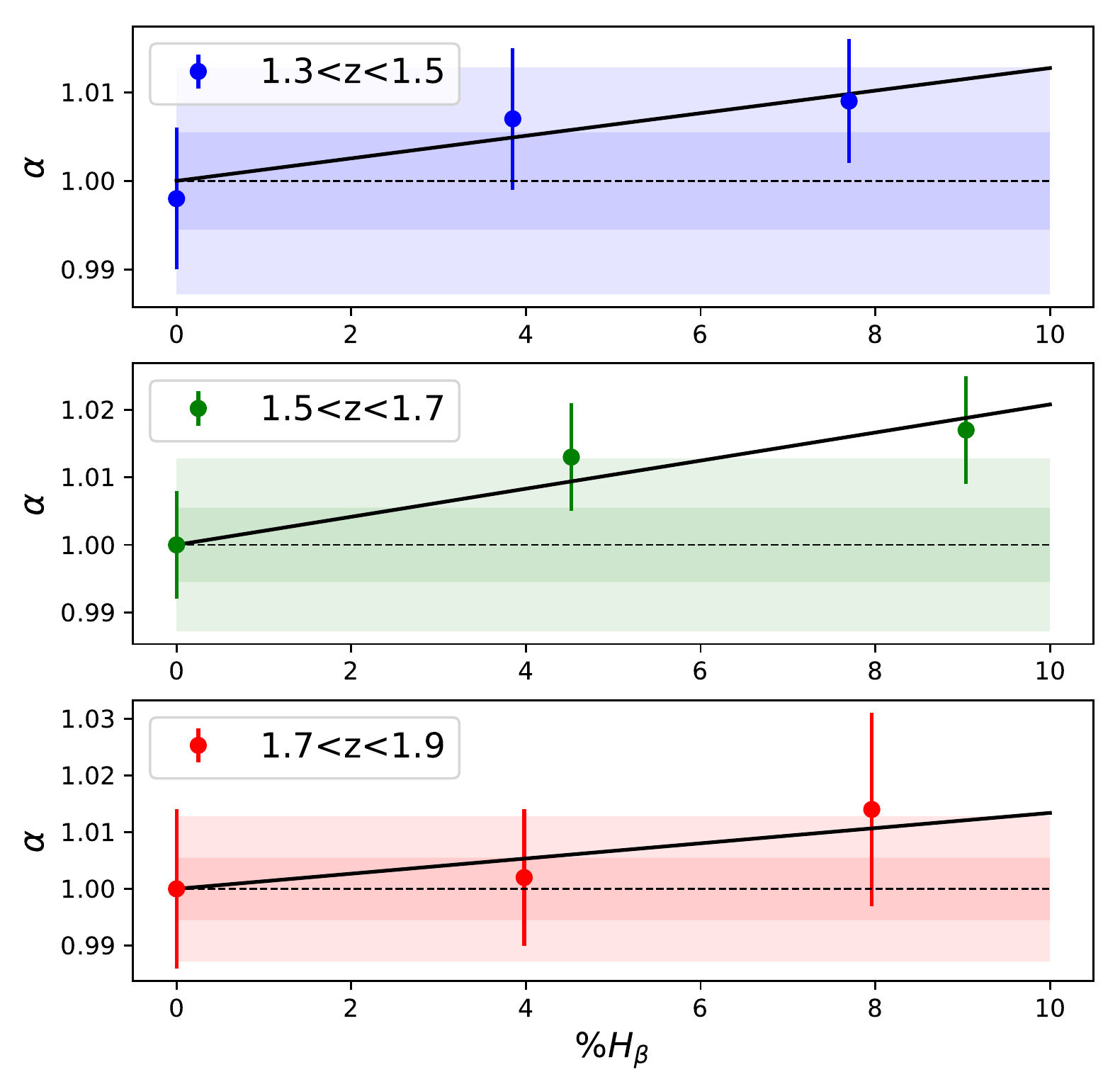}
    \caption{Results for the isotropic parameter $\alpha$ from the BAO fit in real space. $\% H_{\beta}$ indicates the percentage of interlopers over the total number of galaxies. $\% H_{\beta} = 0$ correspond to no interlopers and the $\% H_{\beta} >0$ values correspond to half and the complete number of interlopers in the mocks, for each redshift bin. The covariance matrix has been computed from the 50 mocks.}
    \label{fig:alpha_realSpace_mc}
\end{figure}

\begin{figure}
\includegraphics[width=\columnwidth]{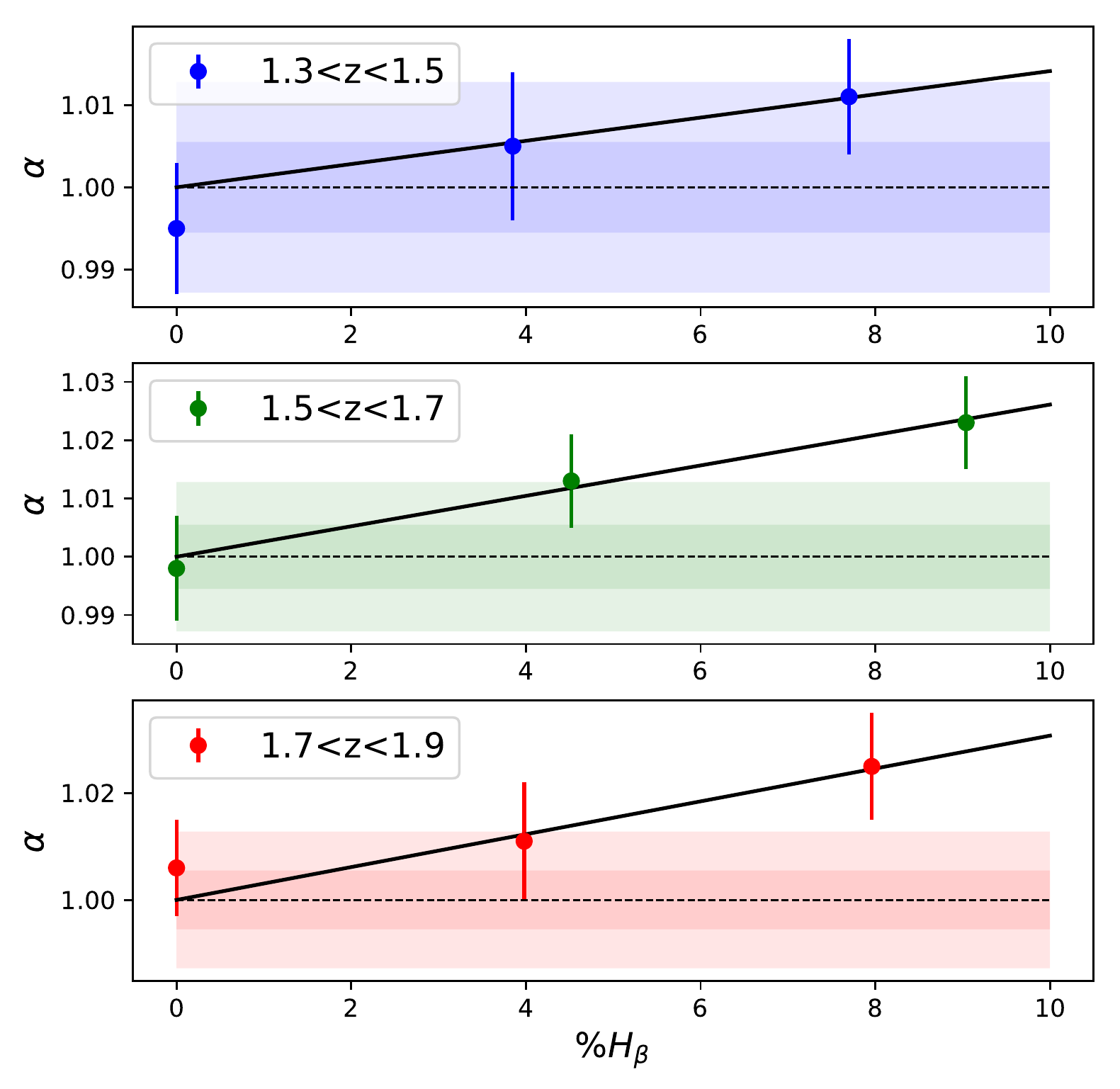}
    \caption{Results for the isotropic parameter $\alpha$ from the BAO fit in redshift space with covariance from the 50 mocks.}
    \label{fig:alpha_RSD_mc}
\end{figure}

We perform the BAO analysis with covariance matrix computed from 50 galaxy mocks, as in equation~\ref{eq:cov_mock}. These mocks are not generated from 50 independent realizations of the dark matter density field and are therefore correlated, so the covariance matrix is not expected to be accurate. Here we use it only to test how the shift in $\alpha$ depends on the covariance matrix used to perform the BAO fit. To this aim, we redo the BAO analysis with this covariance matrix from mocks. The results are Table~\ref{tab:BAO_fit_mc} and Figs.~\ref{fig:alpha_realSpace_mc}, and \ref{fig:alpha_RSD_mc}. 

We perform a linear fit to model the parameter $\alpha$ as a function of the percentage of interlopers $\%H\beta$ over the total number of galaxies, as in Section~\ref{sec:BAOfit}. The best values for the slope parameter in different redshift bins are shown in Table~\ref{tab:b_fit_mc}.

The best fit values of $\alpha$ are very similar to the ones obtained with the Gaussian covariance matrix in Section~\ref{sec:BAOfit}, although their errors are bigger when using the covariance from the mocks. The slopes of the linear fit are also similar in the two cases.

\begin{table}
\begin{center}
\begin{tabular}{| l | c | c | c |}
\hline
Case & $1.3 < z < 1.5$ & $1.5 < z < 1.7$ & $1.7 < z < 1.9$ \\
\hline
\hline
Real: $f_{H\beta} = 0$ & $0.998 \pm 0.008$ & $1.000 \pm 0.008$ & $1.000 \pm 0.014$\\
\hline
Real: $f_{H\beta} = 0.5$ & $1.006 \pm 0.008$ & $1.013 \pm 0.008$ & $1.002 \pm 0.012$\\
\hline
Real: $f_{H\beta} = 1$ & $1.009 \pm 0.007$ & $1.017 \pm 0.008$ & $1.014 \pm 0.012$\\
\hline
\hline
RSD: $f_{H\beta} = 0$ &  $0.995 \pm 0.008$ & $0.998 \pm 0.009$ & $1.006 \pm 0.009$\\
\hline
RSD: $f_{H\beta} = 0.5$ & $1.005 \pm 0.009$ & $1.013 \pm 0.008$ & $1.011 \pm 0.011$\\
\hline
RSD: $f_{H\beta} = 1$ & $1.011 \pm 0.007$ & $1.023 \pm 0.008$ & $1.025 \pm 0.010$\\
\hline
\hline
\end{tabular}
\caption{\label{tab:BAO_fit_mc}Values of the isotropic dilation parameter $\alpha$ from the BAO fit in real and redshift space using the covariance matrix from 50 mocks. $f_{H\beta}$ indicates the fraction of interlopers with respect to the total amount present in the \oiii+H$\beta$ catalogs.}
\end{center}
\end{table}

\begin{table}
\begin{center}
\begin{tabular}{| l | c |}
\hline
Space \& redshift & b slope \\
\hline
\hline
Real: $1.3 < z < 1.5$ & $(1.27 \pm 0.59) \cdot 10^{-3}$\\
\hline
Real: $1.5 < z < 1.7$ & $(2.08 \pm 0.56) \cdot 10^{-3}$\\
\hline
Real: $1.7 < z < 1.9$ & $(1.34 \pm 1.23) \cdot 10^{-3}$\\
\hline
\hline
RSD: $1.3 < z < 1.5$ & $(1.41 \pm 0.60) \cdot 10^{-3}$\\
\hline
RSD: $1.5 < z < 1.7$ & $ (2.61 \pm 0.56) \cdot 10^{-3}$\\
\hline
RSD: $1.7 < z < 1.9$ & $(3.08 \pm 0.81) \cdot 10^{-3}$\\
\hline
\hline
\end{tabular}
\caption{\label{tab:b_fit_mc} Results for the linear fit $y = 1 + {\rm b}\,x $, where $y$ is the isotropic parameter $\alpha$ and $x$ is the percentage of interlopers $\%H\beta$, in real and redshift space.}
\end{center}
\end{table}


\section{\oiii\ and H$\beta$ lines in MOSDEF}
\label{sec:mosdef}
\begin{figure}	\includegraphics[width=\columnwidth]{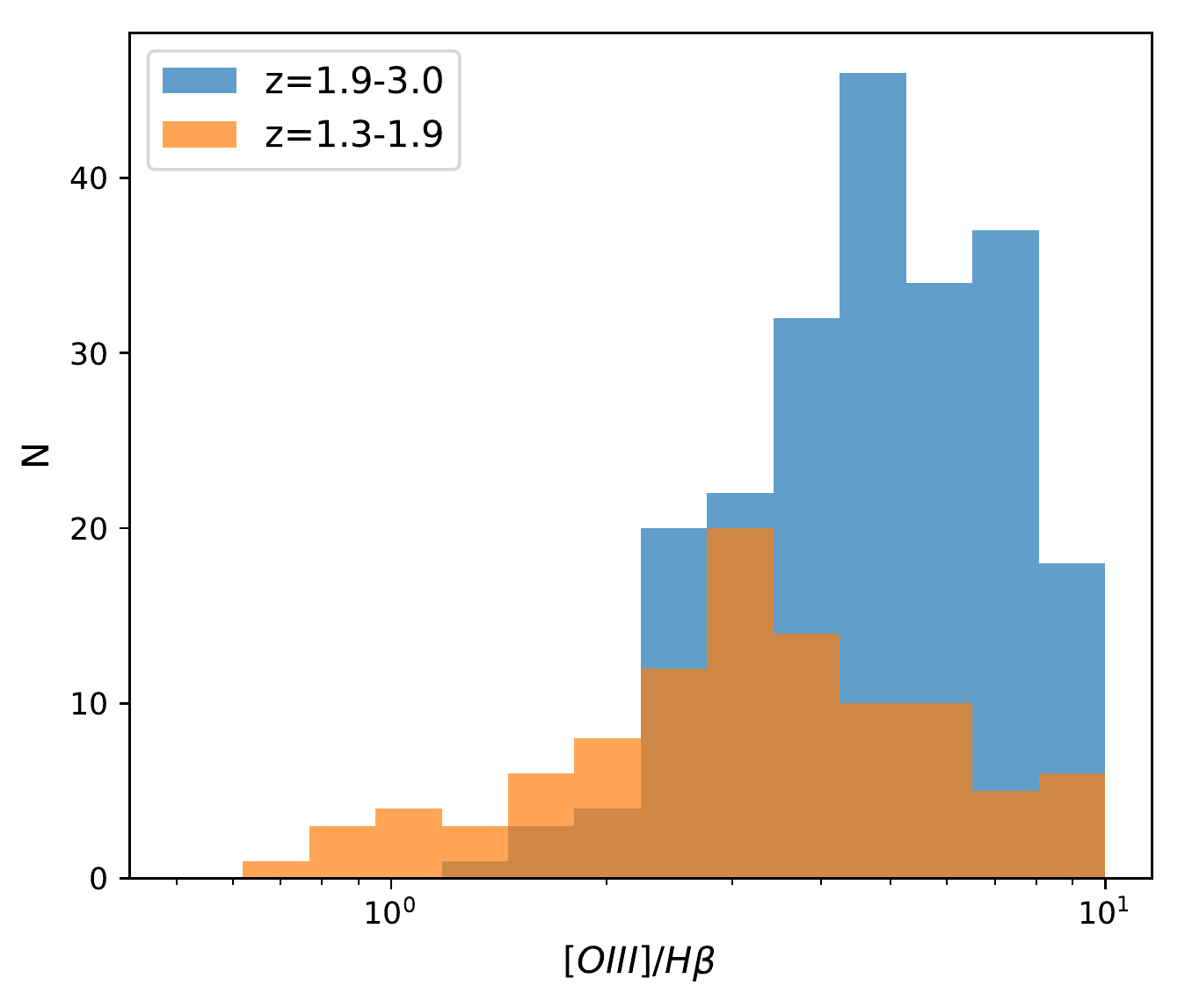}
    \caption{\oiii /H$\beta$ equivalent width ratio distribution in the MOSDEF survey, at redshifts $1.3<z<1.9$ (orange) and $1.9<z<3.0$ (blue).}
    \label{fig:mosdef_EWratio}
\end{figure}

In this section we present relevant results obtained using the data from  \citep{2018ApJ...869...92R} and collected by the MOSDEF survey. Fig.~\ref{fig:mosdef_EWratio} shows the distribution of \oiii\ to H$\beta$ equivalent width ratio for galaxies in redshift bins $1.3<z<1.9$ (orange) and $1.9<z<3.0$ (blue). The distribution in the range $1.3<z<1.9$ presents a peak around the value $3$. In the Roman-like catalog generated for this work, the same PDF peaks around the value $1.5$ (see Fig.~\ref{fig:OIII_Hbeta_ratio}). This indicates that the catalog might contain an overestimated number of interlopers, and the expected interloper fraction could be below $8\%$ within $1.3<z<1.9$. 

\begin{figure}	\includegraphics[width=\columnwidth]{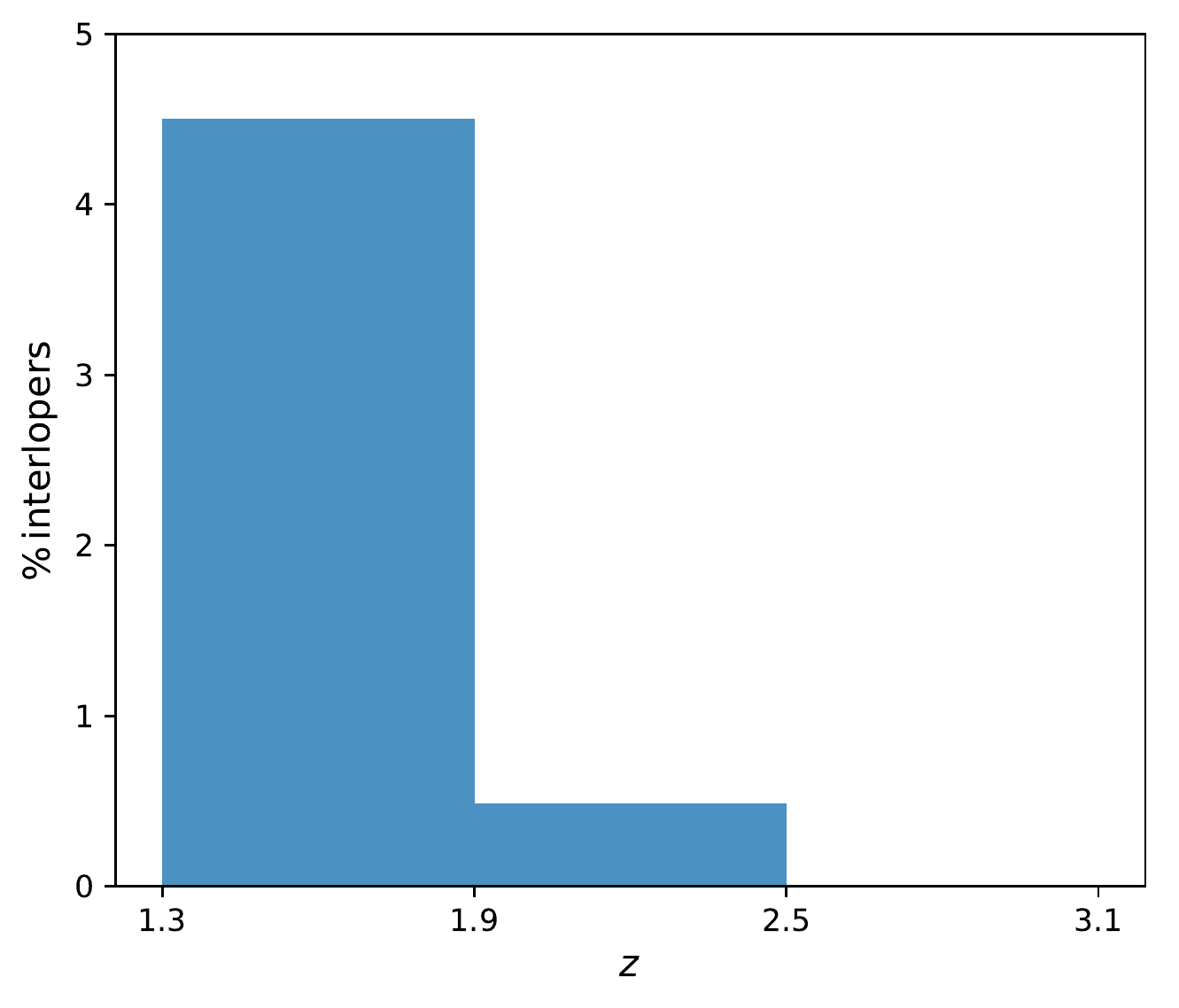}
    \caption{Fraction of H$\beta$ interlopers in the MOSDEF survey at different redshifts.}
    \label{fig:mosdef_interlopers}
\end{figure}

Fig.~\ref{fig:mosdef_interlopers} displays the number of interlopers in the MOSDEF survey. Here interlopers are defined as galaxies with H$\beta$ flux above detection limit (defined as $3\sigma$ limit) and \oiii\ flux below detection limit, while the whole selected galaxies are interlopers or galaxies with \oiii\ flux above detection limit. Under these assumptions, the percentage of interlopers in the sample selected from the MOSDEF survey is $4.5\%$ in redshift range $1.3<z<1.9$, and $0.5\%$ in redshift range $1.9<z<2.5$. Therefore, the galaxies observed in the MOSDEF survey suggest that considering an interloper fraction equal to $8\%$ in Roman is a conservative assumption, and that the percentage of H$\beta$ interlopers should decrease when considering higher redshift ranges. 

\bsp	
\label{lastpage}
\end{document}